\documentclass[
reprint,
iopart
]{revtex4-1}
\usepackage{amsmath} %this allows use of Equation mode
\usepackage[T1]{fontenc}
\usepackage[latin1]{inputenc}
\usepackage{amssymb}
\usepackage{setspace}
\usepackage{color}
\usepackage{float}
\usepackage[english]{babel}
\usepackage[squaren]{SIunits}
\usepackage{dcolumn}% Align table columns on decimal point
\usepackage{bm}% bold math
\usepackage{txfonts}               % free Times-based fonts for text and math
\usepackage{graphicx}   
\begin{document}

\title{Sandpile modelling of dual location fuelling in fusion plasmas}

\author{C. A. Bowie}
\email{craig.bowie@anu.edu.au}
\affiliation{Australian National University, Canberra, ACT 0200, Australia}

\author{M. J. Hole}
\affiliation{Australian National University, Canberra, ACT 0200, Australia}

\begin{abstract}

We modify the Chapman sandpile model (Chapman \textit{et al} \textit{Physical Review Letters} 86, 2814 (2001)) to form comparisons with pellet pacing, which is used to reduce or eliminate ELMs in a fusion plasma. We employ a variation of that model in which a pedestal with feedback is introduced (Bowie and Hole \textit{Phys. Plasmas} 25, 012511 (2018)), which we further modify to provide for dual fuelling - sand is added both at the centre of the sandpile, and near the edge. We observe that when the additional sand is added at the top of the pedestal, MLEs  are largely suppressed. While this suppression comes at a cost by way of reduction in total energy confinement, that reduction is lower than the reduction in MLE size. The trade-off between MLE suppression and reduction in energy confinement depends not only on the amount of extra sand, but also on its precise location relative to the top of the pedestal. We suggest that the approach of constant dual fuelling may be equally applicable to plasmas, and may suggest a strategy for ELM suppression in fusion plasmas. We observe that when the proposed amount of extra sand is added in 'pellets', using frequencies and amounts based on those proposed for ELM suppression for ITER, MLEs are similarly suppressed, although MLEs are not significantly suppressed when the pellet rate does not substantially exceed the MLE frequency. This suggests that pellet injection at the top of the pedestal at small pellet size and high frequency may represent a reasonable physical proxy for our proposed scheme. However, our results suggest that it is not the synchronisation of pellets to ELM frequencies which is the key factor for ELM suppression in this regime, but rather the introduction of additional fuelling at the top of the pedestal.

\end{abstract}

\maketitle

\section{Introduction \label{sec:Introduction}}
% \vspace{-0.25cm}

Nuclear fusion, if it can be effectively controlled, may be critical to our future energy needs. The primary method of seeking to achieve fusion power is via a plasma which is magnetically confined in a torus known as a tokamak. The goal of fusion research is to increase the fusion triple product of temperature, plasma density, and particle confinement time. A step towards this goal, known as H-mode, occurs when the plasma enters into a higher confinement mode, via a mechanism which is not yet fully understood, but which results in the production of a `pedestal' at the edge of the plasma, in which energy confinement rises sharply over a distance of approx 3\% of the toroidal minor radius\cite{Beurskens2009}. However, with H-mode comes a plasma instability known as an edge localised mode, or ELM, which triggers a loss of confinement~\cite{ASDEX1989}. A large ELM may result in a loss of confinement of 5\%~\cite{ASDEX1989}, or from 10-40\% of the pedestal energy~\cite{Beurskens2009} and can cause damage to the first wall of the tokamak\cite{Igitkhanov2013}. For ITER, an upper tolerable limit for ELMs of $\sim$1\% of the pedestal energy has been suggested\cite{Beurskens2009,Zhitlukhin2007}. Controlling ELMs in H-mode is therefore a key objective of fusion plasma research.

Injection of fuel `pellets' has been extensively used as a candidate for ELM control and reduction in fusion plasmas, using pellets to trigger ELMs to increase ELM frequency ($f_{ELM}$), and consequently decrease their maximum size ($W_{ELM}$), on the basis that $f_{ELM}*W_{ELM}=constant$.~\cite{Hong1999,Baylor2005,Baylor2007,Baylor2013,Baylor2015,Lang2004,Lang2014,Lang2015,Pegourie2007,Rhee2012} Pellet size, frequency, and location have all been tested experimentally on ASDEX Upgrade~\cite{Lang2004,Lang2015, Lang2018}, DIII-D~\cite{Baylor2005,Baylor2013}, JET~\cite{Baylor2015, Lang2011, Lang2015}, and EAST~\cite{Li2014,Yao2017} and ELM control using pellets is being considered for use in ITER~\cite{Doyle2007,Baylor2015}.

Injection of pellets to the top of the pedestal has been suggested to produce ELM pacing with reduced energy loss in modelling by Hayashi~\cite{Hayashi2013}, using the code TOPICS-IB. That modelling suggested that pellets with $\sim$1\% of the pedestal particle content, with speed sufficient to reach the pedestal top, will reduce energy loss significantly. The penetration depth of the pellet depends both on its size and speed, as smaller pellets do not penetrate as far into the plasma before ablation. Experiments at JET determined a minimum threshold pellet size which was necessary to reach the top of the pedestal in order to trigger ELMs~\cite{Lang2011}, where the pellet frequency exceeded the natural ELM frequency. For example, Lang\cite{Lang2015} discusses the use of pellets of $1.5-3.7\times10^{20}$D, introduced into a plasma with particle inventory of $6\times10^{20}$D, i.e. $25-60\%$ of the total plasma inventory. It has also been observed that in a 2016 series of discharges in JET, the highest fusion performance was observed using a particle fuelling scheme consisting of pellet injection combined with low gas puffing~\cite{Kim2018}. Lang \cite{Lang2015} discussed pellets added at lower frequencies (higher $\Delta t_P$) with pellet timing aligned to ELM onset. These pellets triggered ELMs. Lang\cite{Lang2015} observes that as pellets increase the plasma density, this in turn increases the L-H threshold. At DIII-D, pellet injection has been observed to trigger synchronous ELMs with a frequency of $12$ times the natural $f_{ELM}$\cite{Huijsmans2015,Baylor2013}. It is proposed that a dual pellet injection system will be used in ITER with large pellets to provide fuelling, and smaller pellets to trigger ELMs\cite{Baylor2015}, and it has been suggested that a pellet frequency of $\sim45$ times the natural $f_{ELM}$ will be required to provide the necessary reduction in heat flux.

One way of understanding the impact of pellet injection on both confinement and ELM behaviour is to seek to identify a physical system whose relaxation processes have characteristics similar to those of the ELMing process under consideration. Of particular interest is the sandpile~\cite{Bak1987}, whose relevance to fusion plasmas is well known~\cite{Chapman1999,Dendy1997}. 

Sandpile models generate avalanches, which may be internal or result in loss of particles from the system. These avalanches are the response to steady fuelling of a system which relaxes through coupled near-neighbour transport events that occur whenever a critical gradient is locally exceeded. The possibility that, in some circumstances, ELMing may resemble avalanching was raised~\cite{Chapman2001A} in studies of the specific sandpile model of Chapman~\cite{Chapman2000}. This simple one-dimensional N-cell sandpile model~\cite{Chapman2000,Chapman2001A} incorporates other established models~\cite{Bak1987,Dendy1998A} as limiting cases. It is centrally fuelled at cell $n = 1/500$, and its distinctive feature is the rule for local redistribution of sand near a cell (say at $n = k$) at which the critical gradient $Z_{c}$ is exceeded. The sandpile is conservatively flattened around the unstable cell over a fast redistribution lengthscale $L_{f}$, which spans the cells $n = k - (L_{f}  - 1), k - (L_{f} - 2), ... , k+1$, so that the total amount of sand in the fluidization region before and after the flattening is unchanged. Because the value at cell $n = k+1$ prior to the redistribution is lower than the value of the cells behind it (at $n<k+1$), the redistribution results in the relocation of sand from the fluidization region, to the cell at $n = k + 1$. If redistributions are sequentially triggered outwards across neighbouring cells, leading to sand ultimately being output at the edge of the sandpile, an avalanche is said to have occurred. The sandpile is then fuelled again, after the sandpile has iterated to stability so that sand ceases to escape from the system.  

The lengthscale $L_{f}$, normalized to the system scale $N$, is typically ~\cite{Chapman1999,Chapman2001A,Chapman2001B,Chapman2003,Chapman2004} treated as the model's primary control parameter $L_{f}/N$, which governs different regimes of avalanche statistics and system dynamics. Here, we employ a modification to the classic model in which the lengthscale is variable over a distance from the edge, which itself depends upon the energy of the system~\cite{Bowie2018}. As $L_f$ reduces near the edge, the gradient increases at the edge, resulting in a pedestal which is subject to feedback due to the dependence of the distance on the energy. The resulting pedestal was introduced as a proxy for the pedestal of a fusion plasma in a H-mode plasma~\cite{Bowie2018}. The feedback loops were seen to be analagous to the feedback effects intrinsic to the H-mode pedestal in a fusion plasma~\cite{Bowie2018}. It was suggested that reduction of feedback in the pedestal could result in ELM suppression within a H-mode plasma~\cite{Bowie2018}.

Typically, the model is centrally fuelled only. Here, we introduce a new feature, being dual fuelling, in which the sandpile is constantly fuelled concurrently at two locations, in order to observe the effect on energy confinement and mass loss event (MLE) size. We observe that by adding $\sim$2.5\% of the sand at a location near the top of the pedestal (near the edge of the plasma), the maximum amount of sand lost in an MLE ($\Delta S_{max}$) is significantly reduced.

%Unlike some implementations of the Chapman model~\cite{Chapman1999,Chapman2001A,Chapman2001B,Chapman2003,Chapman2004}, but following ~\ref{Bowie2016} and \ref{Bowie2018}, $Z_{c}$ is single valued, rather than being randomized. The phenomenology generated by this model has several features resembling tokamak plasmas, including edge pedestals, enhanced confinement~\cite{Chapman2001A} and self-generated internal transport barriers~\cite{Chapman2003}. Particularly relevant here are the systemwide avalanches, or MLEs, resulting (unlike the more numerous internal avalanches which are not considered here) in mass loss from the sandpile.

\section{Dual-fuelled sandpile}

We begin with a feedback model in which sand is added only at the core (as is typical for other implementations of the model). We add sand at a constant rate ($dx_{fc}=1.2$) until the sandpile builds up and enters a `steady state' in which the time averaged amount of sand lost in MLEs equals the amount of sand added. The median waiting time, $\Delta t_n$, between MLEs is  $\sim$$135000$, and $\Delta S_{max}$ is  $\sim$$630000$. The energy of the system ($E_p$), measured by the sum of the squares of the values of the cells, is $\sim$$2.7\times10^9$. The parameters chosen are based on Bowie and Hole~\cite{Bowie2018}.

For the sandpile chosen, the width of the pedestal, $P_w$, is $\sim$$15/500$ cells, meaning that the top of the pedestal is located at $n=485$. Due to the feedback effects built into the model, the pedestal edge moves with time, approximately synchronously with $E_p$. The resulting shape of the sandpile is shown in Figure \ref{fig:Sandpile, Ep, and Max MLE}(a), with the values of $E_p$ and $P_w$ over 2 million iterations shown in Figure \ref{fig:Sandpile, Ep, and Max MLE}(b) and (c). 

\begin{figure}%
\centering
\begin{minipage}{1.2in}%
\includegraphics[clip,trim=0cm 0 0cm 0,width=\linewidth]{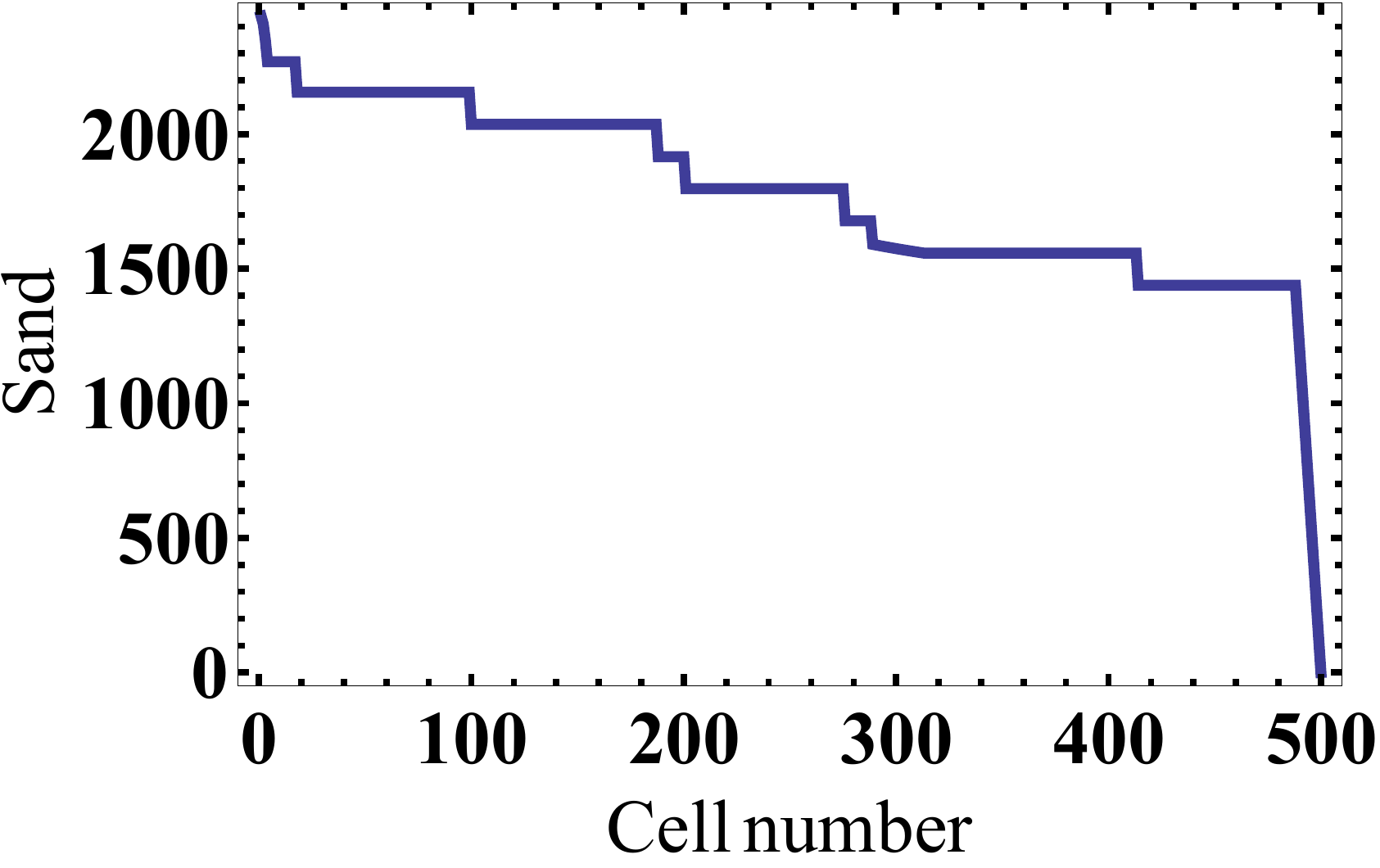}
\end{minipage}%
\begin{minipage}{1.2in}%
\includegraphics[clip,trim=0cm 0 0cm 0,width=\linewidth]{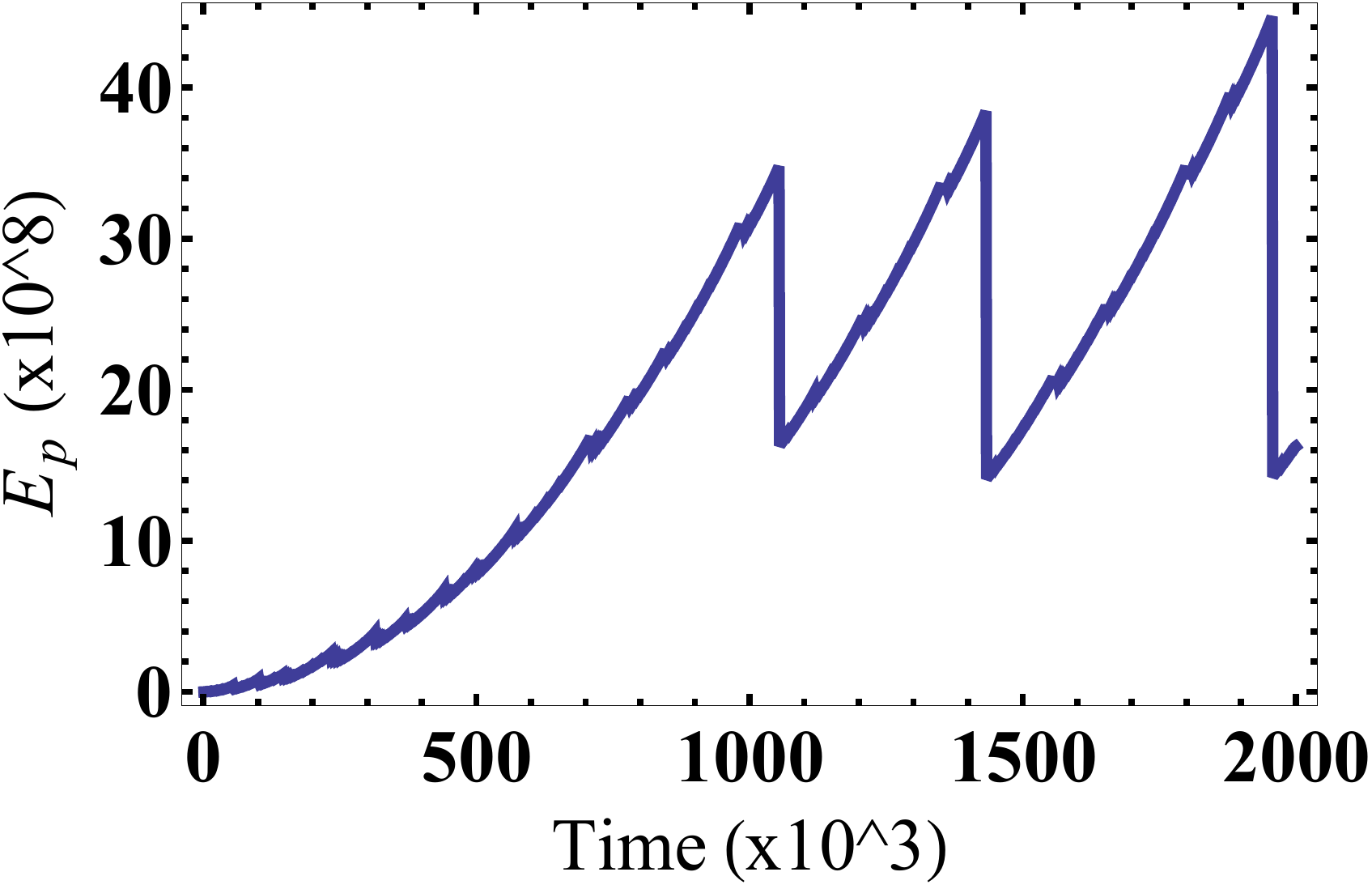}
\end{minipage}%
\begin{minipage}{1.2in}%
\includegraphics[clip,trim=0cm 0 0cm 0,width=\linewidth]{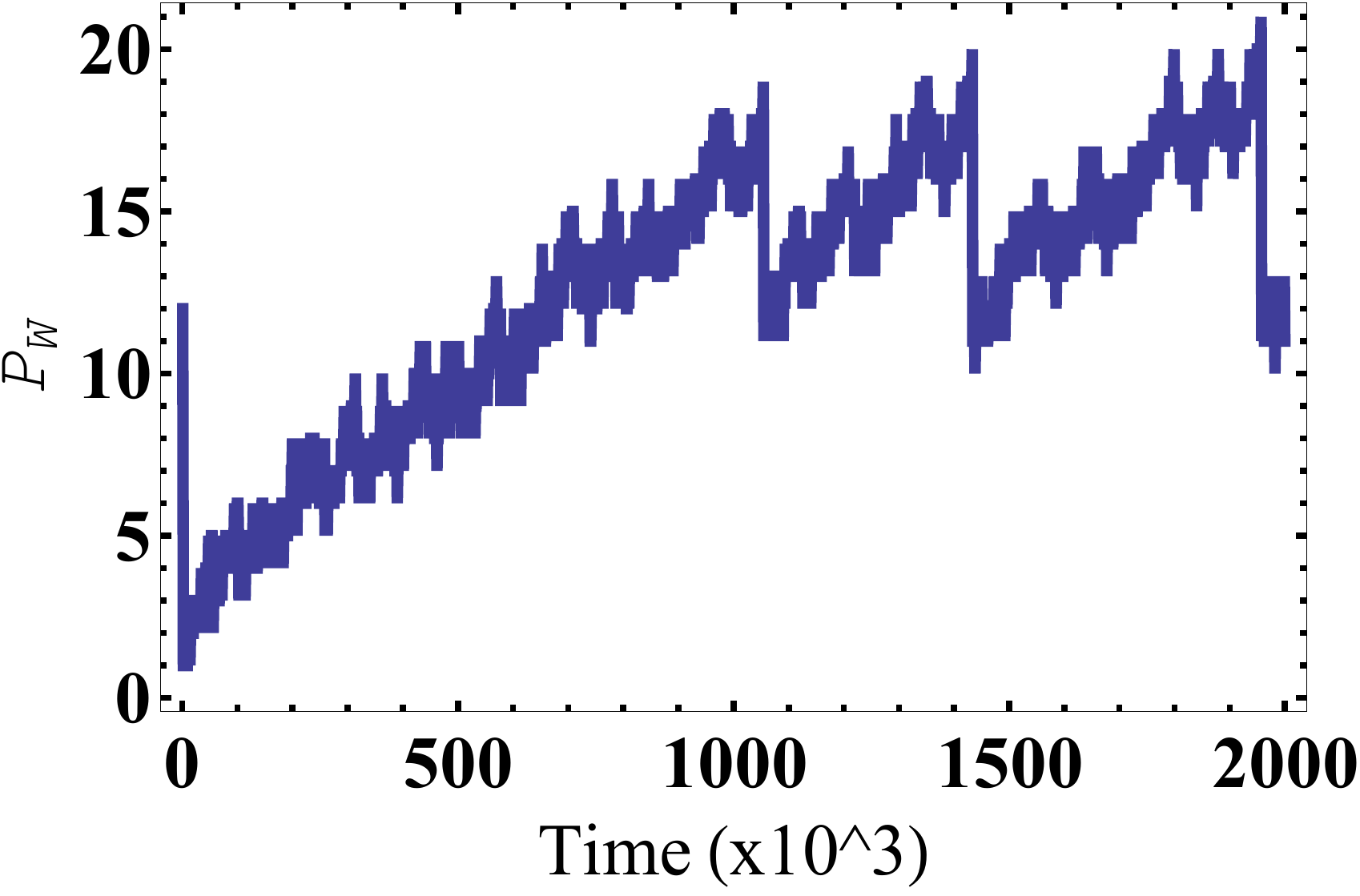}
\end{minipage}

\begin{minipage}{1.2in}%
\includegraphics[width=\linewidth]{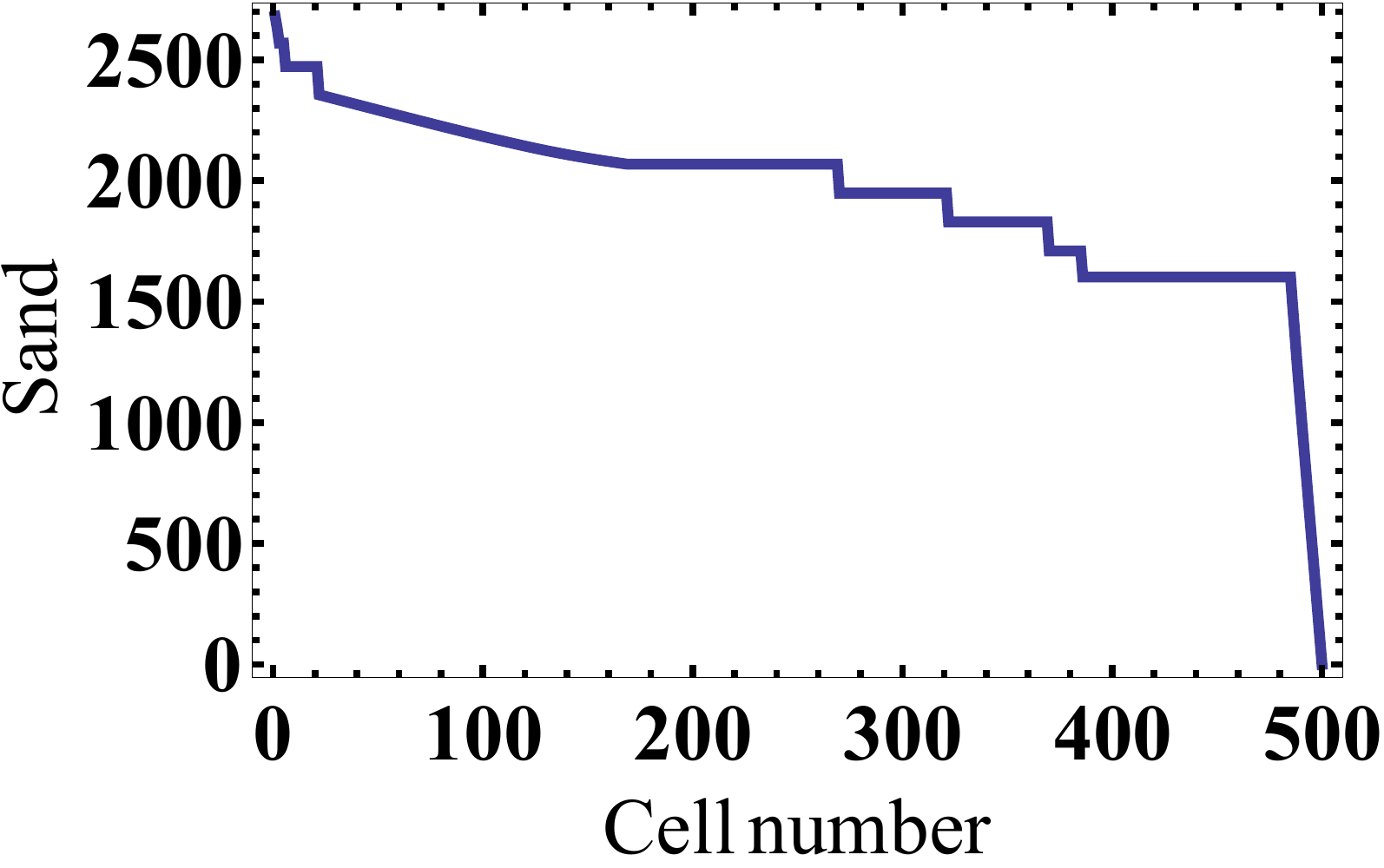}
\end{minipage}%
\begin{minipage}{1.2in}%
\includegraphics[width=\linewidth]{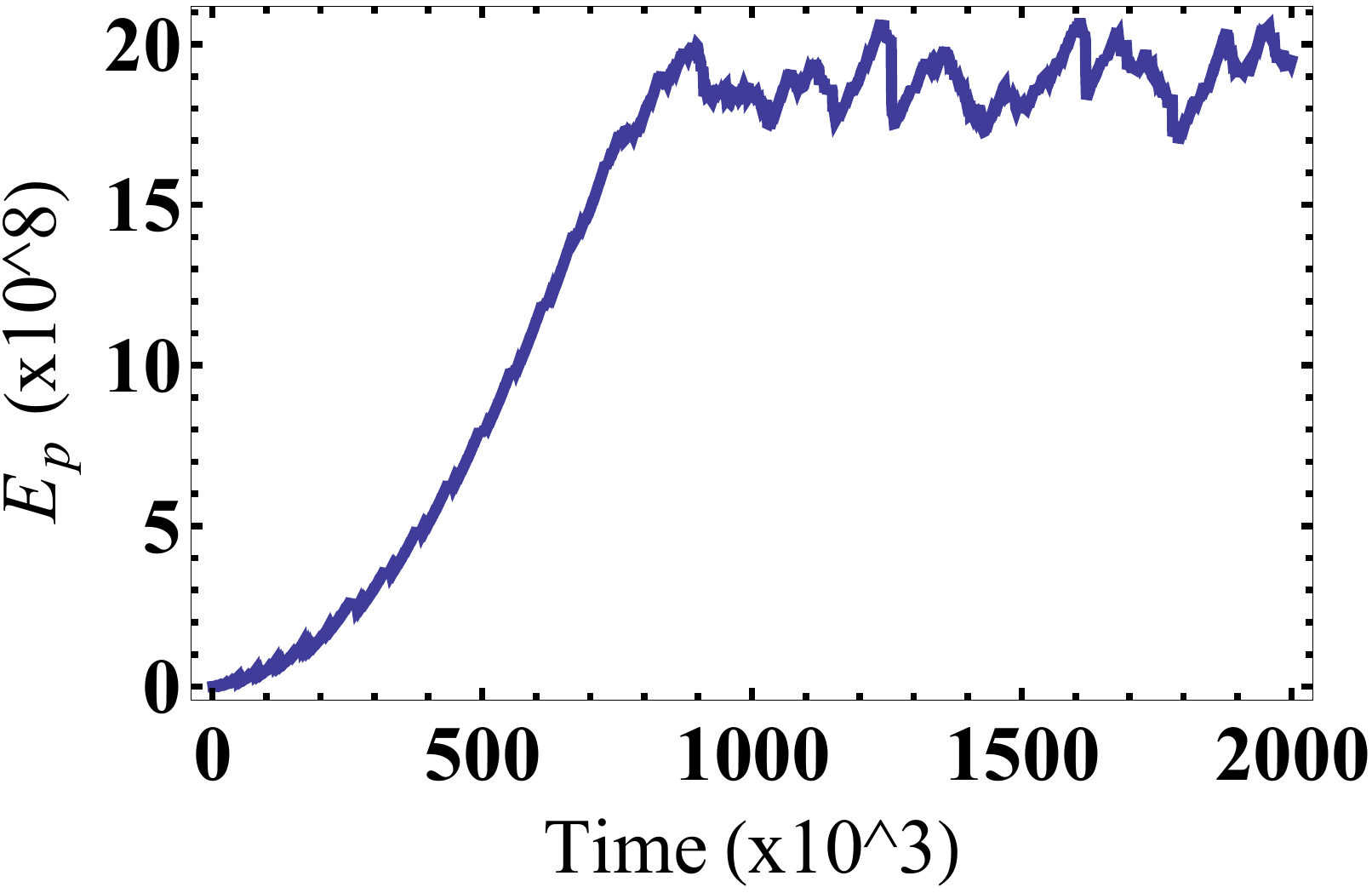}
\end{minipage}%
\begin{minipage}{1.2in}%
\includegraphics[width=\linewidth]{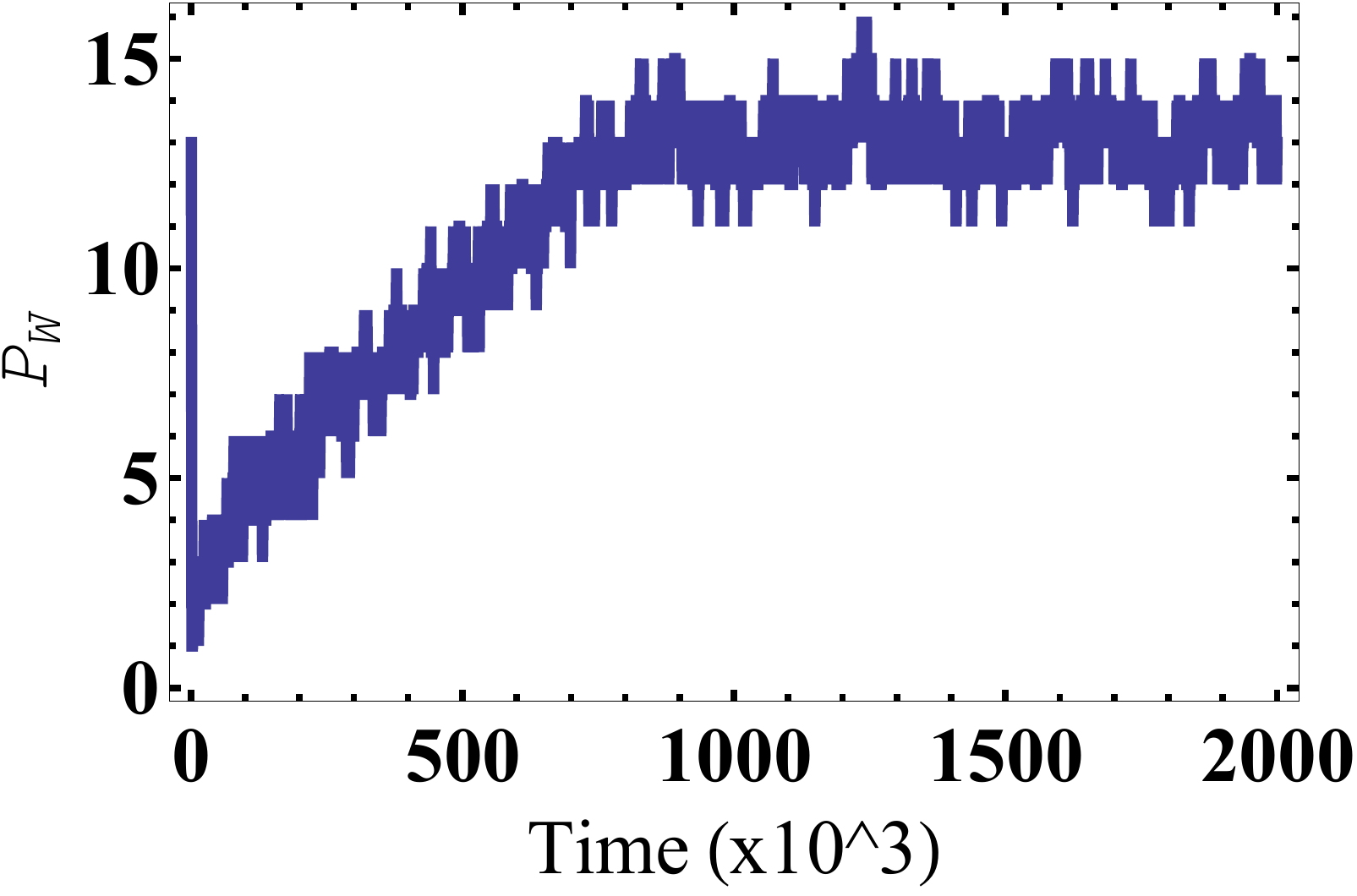}
\end{minipage}%
\caption{(L to R) Sandpile, $E_p$, and $P_w$ plots for base case ($dx_{fe} = 0$) (top); $dx_{fe} = 0.03$, added at $n=487$ (bottom)}%
\label{fig:Sandpile, Ep, and Max MLE}%
\end{figure}

We then modify the model, by adding most of the sand ($dx_{fc}=1.2$) at the core ($n=1$), and some of the sand ($dx_{fe}$)at another location within the sandpile, $n_{fe}$. Sand is added continuously at both the core and $n_{fe}$, representing dual fuelling rather than time separated pellets. We record the average value of $E_p$, and $\Delta S_{max}$. We then repeat the process for a number of values in the range from $n_{fe}=2$ to $n_{fe}=500$. The sandpile, and values of $E_p$ and $P_w$, using this dual fuelling model, are shown in Figure \ref{fig:Sandpile, Ep, and Max MLE}(d-f). We observe that, consistent with the reduction in $\Delta S_{max}$, the $E_p$ and $P_w$ traces are much smoother where dual fuelling is employed. Figure \ref{fig:Sandpile, Ep, and Max MLE}(f) shows us that for $dx_{fe}=0.03$, $P_w$ is about $13$ when $n_{fe}=487$, i.e. the sand is added at about the top of the pedestal.

\begin{figure}%
\centering
\includegraphics[clip,trim={0cm 0cm 0cm 0cm},width=\linewidth]{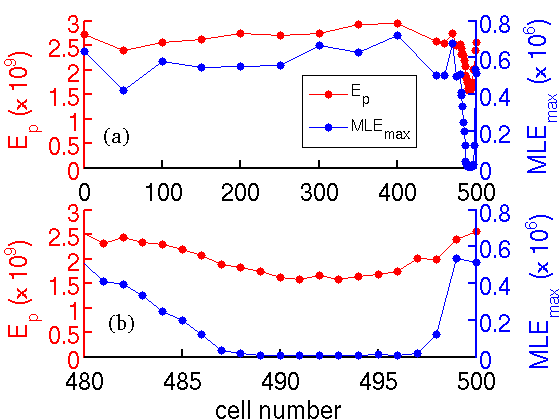}
\caption[Pellet size 0.03]{$E_p$ and $\Delta S_{max}$ for $dx_{fe}= 0.03$, added at $n_{fe}=480$ to $500$ (bottom) and $n_{fe}=1$ to $500$ (top). Figure \ref{fig:pellet-size-0_03}(a) shows the full range from $n=1-500$, while Figure \ref{fig:pellet-size-0_03}(b) shows detail within the pedestal. The average amount of sand in the sandpile is $\sim$$10^6$ units, meaning that $\Delta S_{max}$ is up to $50\%$ for the base case, and is reduced to $\sim$$0.5\%$ when $n_{fe}=487, dx_{fe}=0.03$.}
\label{fig:pellet-size-0_03}
\end{figure}

\begin{figure}
\centering
\includegraphics[clip,trim={0cm 0cm 0cm 0cm},width=\linewidth]{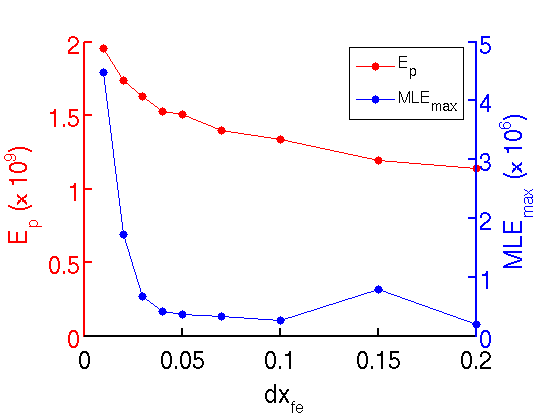}
\caption[Variable pellet sizes]{$E_p$ and $\Delta S_{max}$ for $dx_{fe}= 0.01$ to $0.2$. In each case $n_{fe}=490$. $\Delta S_{max}$ is significantly reduced where $dx_{fe}>0.03$, while $E_p$ declines more slowly, suggesting that $dx_{fe}=0.03$ is optimal for $\Delta S_{max}$, while maintaining $E_p$.}
\label{fig:variable-pellet-sizes---max-mle-and-pe}
\end{figure}

Figure \ref{fig:pellet-size-0_03} shows how $E_p$, and $\Delta S_{max}$ vary as we change $n_{fe}$, for $dx_{fe}=0.03$. Both $E_p$, and $\Delta S_{max}$ are minimised when $n_{fe}$ is located within the pedestal. MLEs are maximally suppressed when $n_{fe}$ is in the range from $487 - 497$, with the maximum $E_p$ in that range at $n_{fe}=487$ (i.e. the top of the pedestal). When $n_{fe}$ is located at the top of the pedestal, $E_p$ declines by about 30\%, with a concurrent $\sim$93\% reduction in $\Delta S_{max}$. If $n_{fe}$ is located just outside the pedestal, a reduction in $\Delta S_{max}$ of $\sim$50\% can be achieved with little effect on $E_p$. By contrast, dual fuelling significantly outside the pedestal has little effect on either $E_p$ or $\Delta S_{max}$, as shown in Figure \ref{fig:pellet-size-0_03}(a).

Essentially, what is observed is that $n_{fe}$, when located at the top of, or within, the pedestal, sets a maximum value for $P_w$, by suppressing further growth of $P_w$. This in turn prevents the sandpile from becoming sufficiently large that it collapses.

The trade-off between reduction in $\Delta S_{max}$ and $E_p$ can also be seen if $dx_{fe}$ is varied. In Figure \ref{fig:variable-pellet-sizes---max-mle-and-pe}, we show $\Delta S_{max}$ and $E_p$ for a range of pellet sizes, added at $n_{fe}=490$, which is near the top of the pedestal. We see that as we increase $dx_{fe}$, $\Delta S_{max}$ and $E_p$ both decline. $\Delta S_{max}$ has been reduced by an order of magnitude at $dx_{fe}=0.03$ and remains relatively steady after that, while $E_p$ continues to decrease as we increase $dx_{fe}$. 

In addition, generally speaking, for values of $dx_{fe}$ below $0.03$, the `dip' in $E_p$ and $\Delta S_{max}$ is smaller, and occurs over a smaller range of values of $n_{fe}$. For higher values, the dip is larger over a $\sim17$ cell range, representing an approximate radial width of $17/500=0.034$ of the plasma. The `sweet spot' appears where the dip is over a wide enough range such that extreme precision in adding $dx_{fe}$ is not required, without resulting in a large decrease in $E_p$.

Taking these factors into account, we suggest that the optimal value for $dx_{fe}$ is about $0.03$, or $2.5\%$ of $dx_{fc}$. As noted above, for $dx_{fe}=0.03$, maximal suppression of MLEs, coupled with minimal reduction in $E_p$, occurs at about $n_{fe}=487$, being the top of the pedestal.

\section{Discussion}

To date, pellet fuelling in fusion plasmas has been aimed at the triggering of an ELM immediately following the introduction of a pellet, so as to increase $f_{ELM}$, and consequently decrease $W_{ELM}$, on the basis that $f_{ELM}\times W_{ELM}=constant$.~\cite{Hong1999,Baylor2005,Baylor2007,Baylor2013,Baylor2015,Lang2004,Lang2014,Lang2015,Pegourie2007,Rhee2012}. Here we suggest a potentially different path to ELM reduction, as the dual fuelling proposed here is constant, rather than pelletized, and therefore does not produce MLEs synchronised with the introduction of additional fuelling. Instead, the constant injection of fuel at or about the top of the pedestal in a feedback modified sandpile, when coupled with the feedback mechanism, triggers MLEs more regularly, but still with a waiting time of at least several thousand time steps.

We observe that MLE suppression does not occur when $n_{fe}$ is significantly outside the pedestal in which feedback occurs. MLE suppression also does not occur for dual fuelling in the classic sandpile model, in which no feedback occurs. This suggests that MLE suppression by dual fuelling is directly related to modification of feedback in the pedestal.

The feedback model, including a pedestal, has been suggested to be analogous to a fusion plasma, including a H-mode pedestal in which feedback effects occur\cite{Bowie2018}, perhaps because a common underlying dynamical behaviour occurs in both the model and the fusion plasma. As a result, we suggest that dual fuelling in a fusion plasma may similarly lead to ELM suppression. Specifically, it may be advantageous to operate a fusion plasma in a mode in which most of the fuelling occurs at the core, while 2.5\% of the fuelling occurs at the top of the pedestal. If our conjecture is correct, and the fuelling properties/insights of the MLE model are portable to a tokamak, such an operating mode will result in the suppression of ELMs at a low energy density and temperature cost.

Notwithstanding that existing pellet fuelling schemes have been aimed at the triggering of an ELM immediately following the introduction of a pellet, there may nonetheless be a relationship between the proposal here and pellet fuelling schemes employed to date. Minimum pellet sizes have been suggested for production of ELMs in experiments, as a consequence of the practical requirement that pellets be large enough to reach the top of the pedestal. The minimum size is also a function of pellet velocity, as the pellet size necessary to reach the top of the pedestal decreases as pellet velocity increases. These minimum sizes are coupled with the maximum practically achievable injection frequency in each experiment. If our analogy is correct, the minimum necessary size to reach the top of the pedestal will couple with the injection frequency to produce an optimal injection frequency, which may be less than the maximum achievable injection frequency.

In order to make a comparison with the proposed ITER scheme, we have 'pelletized' $dx_{fe}$ by adding sand at every $4,000$ time steps (being approximately the natural waiting time in the model, divided by $45$, based on the assumption that the pellet frequency in ITER will be $45$Hz\cite{Baylor2015}), with $f_{ELM}=1$Hz\cite{Baylor2015}. The amount of sand added in total is equal to the amount added continuously, i.e. $4000\times0.03=120$. On the assumption that pellets take effect over their ablation time, rather than instanteously, we have delivered the pellet over $400$ time steps, adopting an observed ablation time for a MAST pellet of $13\times200 \micro \second = 2.6 \milli \second$ \cite{Garzotti2010}, which equates to $\sim 400$ time steps in our model. The result is that at each time step during pellet injection, $dx=1.2$ and $dx_{fe}=0.3$, while for all other time steps $dx=1.2$ and $dx_{fe}=0$. We also observe that the amount of sand in the pedestal in the model is about $11,000$ units, so that a pellet size of $120$ units is $\sim 1\%$ of the particles in the pedestal, which is consistent with modelling by Hayashi\cite{Hayashi2013}, suggesting that the pellet size should be 1\% of the pedestal particle content. With these parameter settings, $E_p\sim1.9\times10^9$ (a reduction of $\sim30\%$ from the base case), and $\Delta S_{max}\sim13000$ (a reduction of $\sim98\%$). 

By contrast, if pellets are injected at a rate equal to the natural MLE frequency, consistent with pellet pacing experiments at JET, then while $E_p\sim1.9\times10^9$ (the same as for the reduction from the base case of $\sim30\%$), $\Delta S_{max}\sim99000$ (a reduction of only $\sim75\%$). The continuing occurrence of significant MLEs is consistent with the result observed at JET in which ELMs still occurred during pellet pacing, rather than being fully suppressed.

This suggests that a series of pellets, such as those to be used in ITER, represent a good approximation to the continuous edge fuelling proposed here, particularly with regard to the practical limitations of implementing such a scheme. Our model also suggests that the relevant consideration for pellet pacing is whether the total amount of particles delivered reaches the ELMing threshold, whether delivered continuously, or over several pellets or gas puffs. This result contrasts with pellet pacing schemes in which pellet timing is aligned to ELM onset \cite{Lang2015} - our result suggests that it is not synchronisation of the pellets which is relevant in this regime, but instead the total amount of fuelling delivered (at least quasi-continuously) at the top of the pedestal.

The scheme may alternatively be implemented by gas puffing, to the extent that gas puffs can be controllably injected at the top of the pedestal as part of a dual fuelling scheme in the proportions suggested here.

\section{Conclusion}

We have implemented a feedback modified sandpile model, to which we have added dual fuelling. The sandpile model incorporates feedback effects within an edge pedestal. We have observed that when additional fuelling is added at the top of the pedestal, MLEs are almost entirely suppressed while $E_p$ is reduced to a lesser extent.

We observe that optimal MLE suppression, with minimal $E_p$ reduction, occurs when edge fuelling represents approximately 2.5\% of core fuelling, and the edge fuelling is added at the top of the pedestal. We conjecture that this MLE suppression results from suppression of feedback in the pedestal of the model. We suggest that a similar scheme employed in a fusion plasma may result in the suppression of ELMs at a low particle density and temperature cost.

We have shown that this scheme is related to a scheme of pellet injection at frequencies up to 45 times the natural $f_{ELM}$ proposed for use in ITER\cite{Baylor2015}, and tested in DIII-D\cite{Baylor2013}, and to a scheme modelled by Hayashi\cite{Hayashi2013},who suggests that small pellets of the order of 1\% of the pedestal particle content, which are fully ablated at the top of the pedestal, may be sufficient to trigger ELMs, and thereby reduce their size. However, significant ELM suppression may not occur unless the pellet rate significantly exceeds $f_{ELM}$. Our result suggests that it is not the synchronisation of pellets to ELMs which is relevant for ELM suppression in this regime, but rather the total amount of fuel delivered (at least quasi-continuously) at the top of the pedestal.

Gas puffing which provides relatively constant edge fuelling may also suppress ELMs at the same ratio of core to edge fuelling.

We suggest that others may wish to implement the scheme proposed here in a fusion plasma, to determine whether edge fuelling can suppress ELMs at a particle density and temperature cost which is considered acceptable for the experiment in question, consistent with the results of our model.

%\vspace{-0.5cm}
\section*{Acknowledgments}
%\vspace{-0.5cm}
%\noindent 
This work was jointly funded by the Australian Research Council through grant FT0991899 and the Australian National University. One of the authors, C. A. Bowie, is supported through an ANU PhD scholarship, an Australian Government Research Training Program (RTP) Scholarship, and an Australian Institute of Nuclear Science and Engineering Postgraduate Research Award.


%\vspace{-1.5cm}
%\bibliographystyle{unrst}
\renewcommand\refname{}
%the command above removes the section heading references
%\bibliography{Dual_fuelling_article_Arxiv}

\end{document}